\documentclass[seceq, epsf]{ptptex}
\usepackage{graphicx}
\usepackage{amsmath,amssymb}
\usepackage{bm}
\usepackage{latexsym}

\title{
Space-Time Uncertainty and Approaches to 
D-Brane Field Theory%
}

\author{
Tamiaki \textsc{Yoneya}%
}

\inst{
Institute of Physics,  University of Tokyo 
\\
 Komaba, Meguro-ku, Tokyo 153-8902
}

\abst{
In connection with the space-time uncertainty principle which 
gives a simple qualitative characterization of non-local or 
non-commutative nature of short-distance space-time structure 
in string theory, author's recent 
approaches toward field theories 
for D-branes are briefly outlined, putting emphasis on some key 
ideas lying in the background.  The final section of 
the present report is devoted partially to a tribute to Yukawa on 
the occasion of the centennial of his birth. }

\begin{document}
\newcommand{\Tr}{{\rm Tr}}
 \newcommand{\llbracket}{[\hspace{-1.6pt}[}
\newcommand{\rrbracket}{]\hspace{-1.6pt}]}
\newcommand{\llbracketbar}{[\hspace{-1.5pt}[ \hspace{-7pt}\smallsetminus}
\newcommand{\rrbracketbar}{]\hspace{-1.5pt}]\hspace{-10pt}\smallsetminus}
\newcommand{\vertbar}{\hspace{3pt}|\hspace{-9pt}\smallsetminus\hspace{-3pt}}
\newcommand{\slashldelimit}{\hspace{-3pt}\smallsetminus}
\newcommand{\rhdllbra}{\triangleright\hspace{-8pt}[\hspace{-1.6pt}[ \hspace{4pt}}
\newcommand{\rhdrrbra}{\, \triangleright \hspace{-5pt}\rrbracket \,}

\maketitle

\section{Introduction}

Most of physicists who are exploring possible roads towards 
the unification of General Relativity and Quantum Theory 
would agree that the space-time structure at 
short distances near the Planck scale can never be 
fully understood in terms of traditional Riemannian geometry. 
Apparently, various geometrical notions in Riemannian geometry 
become more and more inappropriate in describing 
quantum theory as we try to probe shorter and shorter distances. 
One of the concrete phenomena exhibiting this is of 
course the non-renormalizability of ultraviolet divergences 
of general relativistic field theories, including supergravity 
which may be regarded as the final and most extreme outcome 
of unified field theories within the framework 
of {\it classical} local field theory. 
One can imagine a multitude of {\it ad hoc} mechanisms 
which might resolve the ultraviolet difficulty, such as 
space-time lattices, non-local field theories, 
space-time quantizations, so on and so forth.  
However, most of such attempts in the past except for 
string theory lead either to inconsistent 
theories which cannot be compatible with unitarity of quantum 
theory, or to uncontrollable theories without reliable principles 
to fix their contents. At least in its known perturbative definitions,  
string theory is quite rigid and controllable, and its S-matrix is 
perfectly consistent 
with general principles of unitarity. 

We should expect that the right approach toward our goal 
would necessarily 
be a unified theory of all interactions including gravity. 
It is remarkable that string theory has been providing an ideal 
perspective on this fundamental question, 
even though the string theory stemmed originally 
from what seemed nothing to do with gravity and unification of forces, 
and is admittedly still at an stage of an incomplete theory for making 
definite observable predictions. 
Its impressive developments of more than three decades 
convince us of the belief that the string theory is exhibiting
 some of key ingredients 
for achieving the ultimate
 unification. It is the author's conviction 
that the string theory is far deeper than what has been understood so far, 
and that there must be many facets yet to be discovered. 
This was basically the author's attitude when he 
proposed the space-time uncertainty relation as a simple 
qualitative characterization of the short-distance space-time 
structure of the string theory, 20 years ago. 

In the present talk, we would like to first revisit the space-time 
uncertainty relation. Then starting from it as a motivation, 
I will go into some of my recent attempts toward 
the quantum field theory for D-branes. It turned out that 
our models for field theory for D-branes, in particular D-particles,  have certain 
features which are not unrelated to the 
old idea of Yukawa, especially, the idea of `elementary domains'. 
In fact, the idea of space-time uncertainty itself has some 
flavor related to this idea. 
So in the final part of the present talk, I would like to devote my discussion 
partially to a homage to Yukawa,  
to an extent that is related to our formulation 
of D-brane field theory.   That seems appropriate in this Nishinomiya-Yukawa symposium, 
especially since this year 2007 is the centennial of his birth. 

\section{Revisiting the space-time uncertainty principle}

Let me allow to quote some sentences from my 
talk \cite{nyukawa} at the 2nd 
Nishinomiya-Yukawa symposium in 1987, 
in which I have given one of the earliest accounts on the idea \cite{yonenishijima}\cite{yonemodpl}
of the space-time uncertainty relation. 

\vspace{0.2cm}
\noindent
{\it``This implies that the very notion of string itself must be a more 
fundamental geometrical entity encompassing the notion of metric 
tensor, connection and curvature and so forth, of Riemannian 
geometry which is the appropriate language for General 
Relativity based on local field theories. Note that the string 
theory by definition assumes that evergthing is made out of 
strings. Even the geometry of space-time must be expressed 
by using only the notion of strings. This seems to require a 
fundamental revision on the usual concept of the 
space-time continuum. Then the constant $\alpha'$ should be 
interpreted as putting certain limitation on the usual 
concept of Riemannian geometry, just as the Planck 
constant $h$ puts a limitation on the classical concept 
of the phase space. From this point of view, we expect 
some clear characteristic relation expressing the limitation, 
in analogy with the commuation relation 
or Heisenberg's uncertainty principle in quantum theory. 

I now would like to suggest a possible hint along this line of 
arguments. Namely, a possible expression of duality in 
space-time languages could be 
\begin{equation}
\Delta t \Delta \ell \gtrsim 4\pi \alpha'
\label{sturelation}
\end{equation}
where $\Delta t$ is an appropriate measure of the 
indeterminacy of propagation length of a string state 
and the $\Delta \ell$ is an appropriate measure of 
the intrinsic extendedness of a propagating string. 
This ``space-time indeterminacy relation" means that 
any observation in string theory probing short ``time" ($\Delta t
\rightarrow 0$) or small ``distance"($\Delta \ell \rightarrow 0$) 
structure in space-time is associated with large indeterminacy 
in the ``dual" variable, $\Delta \ell$ or $\Delta t$, respectively. 
Thus \eqref{sturelation} sets a limitation about the smallness 
of the space-time domain where arbitrary possible observation 
in string theory is performed. 

In the limit of small $\Delta t$, \eqref{sturelation} can in fact be 
derived \cite{yonenishijima} by re-interpreting the Heisenberg 
relation $\Delta t \Delta E \gtrsim h$. \, ....."
}

\vspace{0.2cm}
\noindent
The last statement came from the recognition that 
the typical intrinsic length scale $\ell$ of a string with large energy $E$ 
is $\ell  \sim \alpha'E$ 
(using the natural unit $c=1=h$ here and in what follows
 for simplicity). Namely, because of the
 huge degeneracy 
of string states a large energy given to a string by 
interaction are expensed dominantly in virtual intermediate states 
by exciting higher string modes, rather than by boosting with a 
large center-of-mass momentum, within the constraint of 
momentum conservation. 

 Here I only summarize 
its meanings in connection with recent development of 
string theory related to D-branes. For more detailed 
discussions, I would like to 
refer the reader to the works mentioned above and also to ref. 
\cite{yoneprog2} 
which includes expository accounts from various relevant 
viewpoints. 
\begin{enumerate}
\item[(1)] The relation \eqref{sturelation} gives a 
characterization of world-sheet ({\it i. e.} open-closed and channel) duality 
in terms of simple space-time language. It can also be regarded 
as a consequence of conformal symmetry, especially of 
modular invariance. As such it has a universal applicability 
at least qualitatively 
in understanding string dynamics including D-branes, 
providing that due care is paid for its application 
since precise definition of the uncertainties $\Delta t, \Delta \ell$ 
has not been given, unfortunately at the present stage of 
development. 
\item[(2)] In connection with this, I emphasize that the traditional notion 
of ``minimal length" $\ell \gtrsim \ell_s
\sim \sqrt{\alpha'}$ is  too restrictive to characterize the 
dynamics  of D-branes. For instance, in the case of  D-particle, 
typical scales \cite{dkps} are 
\begin{equation}
\Delta \ell \sim g_s^{1/3}\ell_s, \quad 
\Delta t \sim g_s^{-1/3}\ell_s
\label{scales}
\end{equation}
in the weak-coupling regime $g_s\ll 1$. The spatial scale $g_s^{1/3}\ell_s$ 
is much smaller than the typical string scale $\ell_s$ 
while the time scale is conversely larger, 
in conformity with the space-time uncertainty relation 
\eqref{sturelation}.  
\item[(3)] The relation is consistent with the Yang-Mills description of 
low-energy effective dynamcis of D-branes. For instance, the 
case of D0-branes, D-particles, is described by the 1-dimensional action, 
\begin{equation}
S_{SYM}=\int dt \, {\rm Tr}
\Bigl( {1\over 2g_s\ell_s} D_t X^i D_t X^i + i \theta D_t \theta
+{1 \over 4g_s\ell_s^5} [X^i, X^j]^2 -
{1\over \ell_s^2}\theta^T \Gamma^i [\theta, X^i]\Bigr).
\label{ymaction}
\end{equation}
This action has a scaling symmetry \cite{jy} under
\begin{equation}
X^i\rightarrow \lambda X^i, \quad t\rightarrow \lambda^{-1}t, 
\quad g_s\rightarrow \lambda^{3-p}g_s
\label{stscale}
\end{equation}
which directly gives \eqref{scales}. 
This symmetry, together with the susy condition 
$\lim_{v \rightarrow 0}S_{eff}=0$, 
constrains the form of 2-body effective action in the form
\begin{equation}
S_{eff}=\int dt \Bigl(
{1\over 2g_s\ell_s}v^2 -\sum_{k=0}^{\infty} c_k {v^{2k}\ell_s^{4k-2}
\over r^{4k-1}}
+O(g_s)\Big)
\end{equation}
and hence effectively governs some of important gross features of D-particle scattering. 
If we re-scale the unit 
globally, the transformation \eqref{stscale} is equivalent to 
the light-cone scaling symmetry in the context of 
the so-called DLCQ M(atrix)-theory \cite{bfss} 
$
t\sim x^+ \rightarrow \lambda^{-2}x^+, \quad X^i 
\rightarrow X^i, \quad x^-\sim R_{11}\, (\equiv g_s\ell_s)
\rightarrow \lambda^2 R_{11}.
$
\end{enumerate}

An important question is what the proper mathematical 
formulation of the space-time uncertainty relation should be. 
Most of discussions at early stages have been 
within the framework of perturbation theory based on the 
world-sheet picture for strings. An obvious possibility 
which does not directly rely on the world-sheet picture  
would be to assume some non-commuative structure for 
the space-time coordinates. One such example is 
 to start from an algebra like $[X^{\mu}, X^{\nu}]^2 \sim \ell_s^4$ 
leading us to an infinite-dimensional 
matrix model which is quite akin to the so-called 
type IIB (IKKT) matrix model, as discussed in detain 
in ref. \cite{yoneprog1}. However, in such approaches, 
 meanings, both physically and mathematically, of the space-time coordinate `operators' $X^{\mu}$ as matrices are quite obscure: 
What are they coordinates of? Are they D-instantons? But then 
how are they related to physical observables? Or are they 
space-time {\it itself}, as for instance in the sense 
that has been argued in \cite{kawaigroup}? ....
In any case, the space-time uncertainty relation in string theory cannot be the primordial principle and 
should be regarded at best as a rough and qualitative consequence resulting from some 
unknown but deeper principles which govern the theory. 
 
Here I would like to recall that the success of quantum field 
theory in the previous century taught us that 
the notion of fields is more important than 
the coordinates, in dealing with various physical phenomena. 
When we make any physical observations, we do not measure 
the coordinates of physical objects directly, but rather effects or events caused by 
the objects. The geometry of space-time must then be expressed 
in terms of fields defined on space-time coordinates 
which themselves have no observational meanings directly 
other than as labels for events. 
Of course,  General Relativity  itself where 
every physical observables can be expressed in terms of fields is a 
realization of this general idea.  
 String field theory is also an attempt 
to realize this\footnote{It is well known that 
a similar viewpoint, `pointless geometry', has been 
emphasized in \cite{wittensft}. It might also be useful here to remind 
that the original conjecture \cite{yonetoyama} by the present author 
of purely cubic
 actions \cite{cubicaction} for string field theory was motivated by this line of thought. } 
in string theory. 
Moreover, the notion of quantized fields 
gave above all the final reconciliation between wave and particle, the primordial duality lying 
in the foundation of modern quantum physics.  

In string theory 
we now 
have understood through developments during recent 10 years that D-branes \cite{pol} can 
be regarded as more basic elements of string theory, rather than 
the (`fundamental') strings.  
One question arising  
then is what are D-branes from this viewpoint. Are they particle-like 
or wave-like objects? All previous discussions of D-branes are actually 
based on the former particle-like view on D-branes: if we go back to the 
effective Yang-Mills descriptions such as \eqref{ymaction}, 
they are obviously configuration-space formulations, 
in the sense that we deal directly with the coordinate matrices $X^{\mu}$ 
whose diagonal components are nothing but the transverse 
positions of D-branes. The Yang-Mills description is an 
approximation ignoring massive open-string degrees of freedom associated to 
D-branes. Thus, open-string field theoies with Chan-Paton 
gauge symmetry are still 
 1st-quantized formulations of D-branes, even though 
they are 2nd-quantized for open strings. Open-string fields 
are in this sense nothing more than the collective 
coordinates for D-branes. To my knowledge, truly 
2nd-quantized theory of D-branes has never been discussed in the literature. 
 I shall explain mainly
 my recent attempts towards 
field theory of D-branes, and
 argue that successful D-brane field theories 
would probably provide us a new perspective on the duality between open and closed 
strings, which is a basic principle  in the heart of various unifications achieved by string theory.

\section{Why D-brane field theory? }
Even apart from the above line of thought, whether {\it field} theory for D-branes 
is possible or not is an interesting question {\it per se}. One motivation 
for D-brane field theory can be explained by making an analogy with 
the well-known Coleman-Mandelstam duality \cite{colman} between 
massive Thirring model and sine-Gordon model 
in two-dimensional field theory. The kink-like soliton excitations 
in the latter can be described in the former as elementary 
excitations corresponding to Dirac fields. This can be established 
explicitly by constructing operators which create and 
annihilate the kink solitons in the latter. D-branes might be understood as  analog 
of the kinks of the sine-Gordon model, since D-branes 
in general appear as non-trivial classical solutions 
(with or without sources) in the low-energy supergravity 
approximation to closed-string field theories.  For string theory, the Schwinger model 
also provides an interesting analogy with open/closed string 
duality: the one-loop of massless Dirac fields gives a pole 
singularity corresponding to massive scalar excitation. 
The massive scalar (or sine-Gordon) field is the analog of closed strings, 
while the massless Dirac (or massive Thirring) field is that of 
open strings.  

\vspace{0.2cm}

\begin{figure}[htbp]
\begin{center}
\includegraphics[width=11cm]{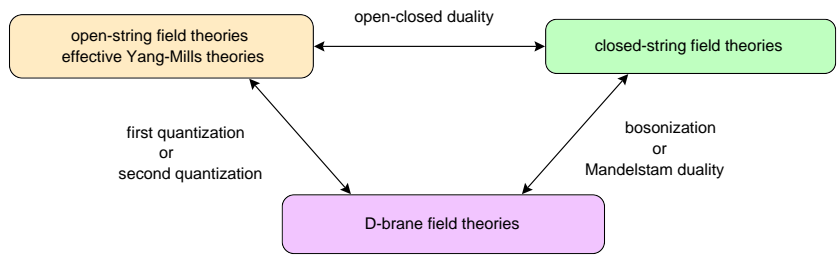}
\caption{\footnotesize{
A trinity of dualities: D-brane field theories point toward 
the third possible formulation of string theory treating 
D-branes as elementary excitations. 
}}
\end{center}
\end{figure}

These analogies suggest the possibility of some 
field theory for D-branes which is dual to closed-string field theory, 
in a similar way as the massive Thirring model is dual 
to the sine-Gordon model. 
The field theory of D-branes must give a 2nd quantized 
formulation of multi-body D-brane systems which in the non-relativistic low-energy approximation should be equivalent 
to the Yang-Mills theory of D-branes. Hence, we should  be 
able to find some framework for 2nd-quantizing the Yang-Mills 
theory in the sense of D-branes. Namely, the number of D-branes corresponds 
to the size of the Yang-Mills gauge groups. Therefore, 
D-brane field operators must act as operators which change the 
size of the gauge groups in the Fock space of D-branes. 

The idea can be summarized by 
the diagram in Fig. 1. 
We expect the existence of 
third possible formulation of string (and/or M) theory. 
D-brane field theories would give 
a new bridge which 
provides an explanation of open/closed string duality 
as a special case of the general notion of particle/wave duality. 
Namely, open string field as a particle picture for D-branes is dual to 
D-brane field theory, a wave picture, 
 by 2nd-quantization, and then D-brane field 
theory is dual to closed-string field theory by a generalized 
bosonization, as summarized by the following diagram: 

\vspace{0.2cm}
\begin{center}
D-particle   \hspace{1.8cm}    --- \hspace{1.7cm}  D-wave\\
\vspace{0.1cm}
$\hspace{0.2cm} | $   \hspace{5cm} $|$\\
\vspace{0.1cm}
open string field theory \hspace{0.3cm} --- \hspace{0.3cm}   D-brane field theory \\
\vspace{0.1cm}
$\, \, \setminus \hspace{4.4cm} \slash$\\ \,
\hspace{0.1cm} closed string field theory
\end{center}
\vspace{0.2cm}

\section{Gauge symmetry as a quantum-statistics symmetry}
In attempting the 2nd-quantization of D-brane quantum mechanics, 
one of the most interesting problems is that the permutation symmetry 
which is the basic quantum-statistical 
symmetry governing the ordinary 2nd-quantization 
of many-particle systems must be replaced by the continuous 
gauge group U($N$). Let $X_{ab}^i$ ($a, b, \ldots =1, 2, \ldots, N$) be the 
coordinate matrices of $N$ D-branes with 
$i$ being the transverse spatial directions. Then, the 
gauge symmetry transformation has an adjoint action as
\begin{equation}
X^i_{ab}\rightarrow (UX^iU^{-1})_{ab}, \quad U\in \mbox{U($N$)}.
\end{equation}
The diagonal components $X_{aa}^i$ represent the positions of $N$ 
D-branes, while the off-diagonal components correspond to 
the lowest degrees of freedom of open strings connecting them. 
If we constrain the system such that only the diagonal elements 
survive, the gauge symmetry is reduced to  permutations 
of diagonal elements. It is therefore clear that one of crucial 
ingredients in trying 2nd-quantizing the system is how to deal with 
this feature. In general, D-branes can be treated neither as 
ordinary bosons nor fermions. 

There is, however, one exception 
where we can apply the usual 2nd-quantization for matrix system. 
If there is only one 
matrix degree of freedom as in the case of the old 
$c=1$ matrix model, we can diagonalize the matrix coordinate 
from the outset, and the Hilbert space is then replaced by that of $N$ 
free fermions, moving in a fixed external potential, whose 
coordinates are those eigenvalues. The 2nd-quantization 
is then reduced to that of usual non-relativistic fermions. 
Unfortunately, this method cannot be extended to cases where 
we have to deal with many matrices. One of popular methods in the past 
has been to try to represent such systems by a set of gauge 
invariants.  A typical idea along this line has been to use Wilson-loop 
like operators. However, from the viewpoint of formulating 
precise dynamical theory, putting aside its use 
merely as  convenient observables to probe 
various physical properties, use of Wilson loops or similar 
set of invariants is quite problematic.  The reason is that it is usually 
extremely difficult to separate independent degrees of 
freedom out of the set of infinite number of invariants. 
This difficulty makes almost hopeless any attempt of 
2nd-quantization for matrix systems
 in our sense. 

I shall suggest an entirely new way of dealing with 
the situation, 
by restricting ourselves to the system of D0-branes in the 
Yang-Mills approximation. 
Before proceeding to this, I wish to mention another related work \cite{yonecuntz} of 
myself, in which an attempt extending the method of 
the $c=1$ matrix model was made 
for a description of the extremal correlators for 
a general set of $1/2$-BPS 
operators for D3-branes. This case is very special in that 
we can justifiably use a free-field approximation.\cite{antal} It was shown that 
in this approximation the extremal correlators 
can be expressed in terms of bilinear operators of D3-brane 
fields. The normal modes of the D3-brane fields are 
composite operators of the form $(b_nc_I, b_n^{\dagger}c_I^{\dagger})$, where $b_n, b_n^{\dagger}$ carrying only 
energy index $n$  are usual 
fermion mode operators obeying the standard canonical 
anti-commutation relations, while $c_I, c_I^{\dagger}$ 
obey the  Cuntz algebra \cite{cuntz} in the form
\[
c_{I_1} c_{I_2}^{\dagger}=
\delta_{I_1I_2}, \quad 
\sum_{I=0}^{\infty}c_I^{\dagger} c_I=1
\] 
and carry only internal R-symmetry indices $I$ which denote 
an appropriate basis for the completely symmetric and traceless 
representations of SO(6). 
As such, the D3-brane 
fields  do not satisfy any simple commutation relations. 
Yet the bilinears representing the ordinary gauge invariants and 
acting on physically allowed states 
satisfy the standard (commutative) algebra 

A lesson from this work was that 
D-brane fields cannot be described in any existing canonical framework of quantum field theory.
It seems to be inevitable that some drastic 
extension of the general framework of quantum field theory 
is necessitated in order to carry out the 2nd-quantization 
of D-branes in general. We have to invent a 
new mathematical framework for establishing 
truly 2nd-quantized field theories of D-branes, 
which obey {\it continuous} quantum statistical symmetry. 
This is an entirely new way of counting the degrees of freedoms 
of physical excitations. The `D-brane statistics' is far more alien 
than Bose-Einstein or Fermi-Dirac statistics were.  

\section{An attempt toward D-particle field theory} 
I now describe my attempt toward a quantum field theory 
of D-particles, in the non-relativistic 
Yang-Mills approximation. What is to be done 
is to find an appropriate mathematical structure for dealing with 
the Fock space 
 of Yang-Mills quantum mechanics in terms 
of field operators which change the number of D-particles. 
For each $N$, the configuration space for the wave function 
consists of 
the $N\times N$ matrix coordinate $X_{ab}$. In the present report we only 
explain some of basic ideas, by referring more details to the 
original work \cite{yonedpfield}. Here and in what follows, we will suppress spatial 
indices and also Grassmannian degrees of freedom.

Let us recall the usual 2nd-quantization for ordinary bosons. 
The configuration-space wave functions of $N$ particles are replaced by 
a state vector $|\Psi \rangle$  in a Fock space $
{\cal F}=\sum_N \bigoplus {\cal H}_N$ as 
\[
\Psi(x_1, x_2, \ldots, x_N)
\]
\begin{equation}
\Rightarrow |\Psi\rangle 
=\Big(
\prod_{i=1}^N \int d^dx_i\Big) 
\Psi(x_1, x_2, \ldots, x_N)
\psi^{\dagger}(x_N)\psi^{\dagger}(x_{N-1}) \cdots \psi^{\dagger}(x_1)|0\rangle
\end{equation}
where the field operators define mappings between 
${\cal H}_N$'s with different $N$, 
\[
\psi^{\dagger}(x): {\cal H}_N \rightarrow {\cal H}_{N+1}
, \]
\[
\psi(x): {\cal H}_N\rightarrow {\cal H}_{N-1}.\]
The quantum-statistical symmetry is expressed as 
\begin{equation}
\psi^{\dagger}
(x_{N})\psi^{\dagger}(x_{N-1}) \cdots \psi^{\dagger}(x_1)|0\rangle
=\psi^{\dagger}(x_{P(N)})\psi^{\dagger}(x_{P(N-1)}) \cdots \psi^{\dagger}(x_{P(1)})|0\rangle
\label{usualstatistics} 
\end{equation}
and 
\begin{equation}
\psi(y)\psi^{\dagger}(x_N)\cdots \psi^{\dagger}(x_1)|0\rangle
={1\over (N-1)!}\sum_P \delta^d(y-x_{P(N)})
\psi^{\dagger}(x_{P(N-1)})\ldots 
\ldots \psi^{\dagger}(x_{P(1)})|0\rangle
\label{usualannihilation}
\end{equation}
where the summation is over all different permutations  $P:(12\ldots N)\rightarrow 
(i_1i_2\ldots i_N), \\P(k)=i_{k}$. 
Of course, the whole structure is reformulated 
as a representation theory 
of the canonical commutation relations of the field 
operators acting on the Fock vacuum $|0\rangle
: 
\psi(x)|0\rangle =0$, 
\[
[\psi(x), \psi^{\dagger}(y)]=\delta^d(x-y), \quad 
[\psi(x), \psi(y)]=0=[\psi^{\dagger}(x), 
\psi^{\dagger}(y)].
\]
In particular, the last two commutators represent the 
permutation symmetry, or Bose statistics. 

If we compare this structure with the Fock space of 
D-particles which consists of the Yang-Mills theory of different 
$N$, two crucial features are that 
\vspace{0.2cm}
\begin{enumerate}
\item[(a)] The increase of the degrees of freedom in 
the mapping ${\cal H}_N
\rightarrow {\cal H}_{N+1}$ is
$d_N\equiv 
d(2N+1)=d((N+1)^2-N^2)$, instead of $d=d(N+1)-dN$ which is 
independent of $N$. 
\item[(b)] The statistical symmetry is a continuous group, the 
adjoint representation of  
U($N$), instead of the discrete group of permutations $\{P\}$.
\end{enumerate}
\vspace{0.2cm}
The feature (a) indicates that  D-particle fields which we denote by $\phi^{\pm}[z,\bar{z}; t]$, creating or annihilating 
 a D-particle,  must be 
defined on a base space with an infinite number of 
coordinate components, since $d_N\rightarrow \infty$ as 
$N\rightarrow \infty$. But, if they act on a state with a 
definite number $N$ of D-particles, only the finite 
number, $d_N$, of them must be activated, and 
the remaining ones should be treated as dummy variables.  
In terms of the matrix coordinates, 
we first redefine components of these infinite dimensional 
space as 
\begin{equation}
z^{(b)}_a=X_{ab} =\bar{X}_{ba}  \quad \mbox{for}  \quad b\ge a, 
\end{equation}
which is to be interpreted as the $a$-th component 
of the (complex) coordinates of the $b$-th D-particle.  
The  assumption here is that the field algebra and its 
representation should be set up such that we can effectively ignore 
the components 
$z^{(b)}_a, \bar{z}^{(b)}_a$ with $a>b$ for the 
$b$-th operation in adding 
D-particles. 
 Hence, the matrix variables are embedded into sets of the arrays  
of infinite-dimensional complex vectors $(z_1=x_1+iy_1,z_2=x_2+iy_2, \ldots)$.   Note that the upper indices with braces discriminate 
the D-particles by ordering them, 
whereas the lower indices without brace represent 
the components of the infinite dimensional 
coordinate vector $(z,\bar{z})=\{z_1,\bar{z}_1, z_2, \bar{z}_2, \ldots \}$ for each D-particle. 

 Thus we define creation, $\phi^+[z, \bar{z}]$, 
and annihilation, $\phi^-[z, \bar{z}]$, operators 
on the base space of an infinite-dimensional 
vector space  consisting of $(z_n, \bar{z}_n)$ with $n=1, 2, 3, ...$. 
The process of  creating and annihilating 
a D-particle must be defined conceptually (time being 
suppressed) as
\[
\phi^+ : \, |0\rangle \rightarrow \phi^+[z^{(1)}, \overline{z}^{(1)}]|0\rangle 
\, \rightarrow \phi^+[z^{(2)}, \overline{z}^{(2)}]\phi^+[z^{(1)}, 
\overline{z}^{(1)}]|0\rangle \, \rightarrow \cdots ,
\]
\[\hspace{0.9cm}
\phi^- : \, 
0 
\, \leftarrow|0\rangle \, \leftarrow
\phi^+[z^{(1)}, \overline{z}^{(1)}]|0\rangle \leftarrow 
\phi^+[z^{(2)}, \overline{z}^{(2)}]\phi^+[z^{(1)}, 
\overline{z}^{(1)}]|0\rangle \leftarrow \cdots .
\]
Pictorially this is illustrated in Fig. 2. 
\begin{figure}[htbp]
\begin{center}
\includegraphics[width=11.3cm]{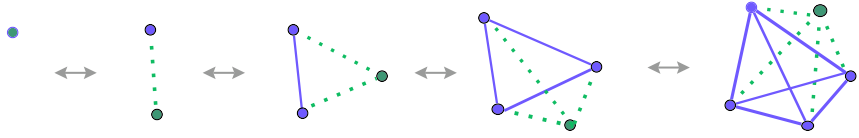}
\caption{\footnotesize{
The D-particle coordinates and the open strings mediating them are 
denoted by blobs and lines connecting them, respectively. 
The real lines are open-string degrees of freedom 
which have been created before the 
latest operation of the creation field operator, while the dotted lines 
indicate those created by the last operation. The arrows 
indicate the operation of 
creation (from left to right) and annihilation 
(from right to left) of D-particles.
}}
\end{center}
\vspace{-0.3cm}
\end{figure}

\vspace{0.2cm}
The presence of the dummy components, the feature (a) above,  is taken into account 
by assuming a set of special projection conditions, such as 
\[
\partial_{y_1^{(1)}}\phi^+[z^{(1)}, \overline{z}^{(1)}]|0\rangle =0, 
\quad 
 \partial_{z^{(1)}_k}\phi^+[z^{(1)}, \overline{z}^{(1)}]|0\rangle =0 \quad 
\mbox{for}\, \quad 
k\ge 2,
\]
\[
\partial_{y_2^{(2)}}\phi^+[z^{(2)}, \overline{z}^{(2)}]\phi^+[z^{(1)}, 
\overline{z}^{(1)}]
|0\rangle =0, 
\, 
 \partial_{z^{(2)}_k}\phi^+[z^{(2)}, \overline{z}^{(2)}]
 \phi^+[z^{(1)}, \overline{z}^{(1)}]|0\rangle =0 \,\,
\mbox{for}\, \,
k\ge 3.
\]
The feature (b), a continuous quantum statistics corresponding to 
gauge invariance, is taken into account by assuming symmetry 
constraints such as
\[
\phi^+[(UXU^{-1})_{12}, (UXU^{-1})_{21}, (UXU^{-1})_{22}]
\phi^+[(UXU^{-1})_{11}]|0\rangle 
\]
\[=
\phi^+[z^{(2)}, \overline{z}^{(2)}]\phi^+[z^{(1)}, \overline{z}^{(1)}]
|0\rangle 
\]
as a natural extension of \eqref{usualstatistics}. The action of the 
annihilation operator is defined as 
\[
\phi^-[z,\overline{z}]|0\rangle =0, 
\]
\[
\phi^-[z, \overline{z}]\phi^+[z^{(1)}, \overline{z}^{(1)}]|0\rangle 
=\delta(x_1-x^{(1)}_1)\delta(y_1)\prod_{k\ge 2}\delta(z_k)|0\rangle, 
\]
\[
\phi^-[z,\overline{z}]\phi^+[z^{(2)}, \overline{z}^{(2)}]
\phi^+[z^{(1)}, \overline{z}^{(1)}]|0\rangle 
\]
\[
=
\int dU \delta^2(z_1-(UXU^{-1})_{12}))\delta(x_2-
(UXU^{-1})_{22})
\Big(\prod_{k\ge 3}\delta(z_k) \Big)\, 
\phi^+[(UXU^{-1}_{11}]|0\rangle\, 
\]
which are again natural extensions of the ordinary one \eqref{usualannihilation} corresponding to the usual discrete 
statistical symmetry. 
Actually, the algebra of these field operators has various 
peculiar features such as non-associativity, and hence we 
need more sophisticated notations to treat this system precisely. 
  
 With this apparatus, it is  possible to represent 
all possible gauge invariants in terms of bilinear operators of the form 
\begin{equation}
\langle \phi^+, F\phi^-\rangle 
\equiv 
\int [d^{2d}z]\, \phi^+[z, \bar{z}] F\phi^-[z, \bar{z}] 
\end{equation}
where $F$ is an appropriate operator acting on arbitrary 
functions on the infinite-dimensional coordinate space of $(z_n, \bar{z}_n)$.   
For instance, the Schr\"{o}dinger equation takes the form
$
{\cal H}|\Psi\rangle =0, 
$ where 
\[
{\cal H}=i(4\langle \phi^+, \phi^-\rangle +1)\partial_t+
\]
\[
2g_s\ell_s\Big(
(\langle \phi^+, \phi^-\rangle +1) \langle \phi^+, 
\partial_{\bar{z}^i}\cdot \partial_{z^i}
\phi^-\rangle 
+3\langle \phi^+, \partial_{\bar{z}^i}\phi^-\rangle\cdot
\langle \phi^+, \partial_{z^i}\phi^-\rangle
\Big)
\]
\begin{equation}
+\frac{1}{2g_s\ell_s^5}
(4\langle \phi^+, \phi^-\rangle +1)(\langle \phi^+, \phi^-\rangle 
+1)\langle \phi^+, \Big(
(\bar{z}^i\cdot z^j)^2 -(\bar{z}^i\cdot z^j)(\bar{z}^j\cdot z^i)
\Big)\phi^-\rangle.
\end{equation}
In the large $N$ limit and in the center-of-mass frame 
satisfying 
\[
\langle \phi^+, \partial_{z^i} \phi^-\rangle =0, 
\]
 this is simplified to 
\begin{equation}
i\partial_t|\Psi\rangle =\Big[
-{g_s\ell_s\over 2}\langle \phi^+, 
\partial_{\bar{z}^i}\cdot \partial_{z^i}
\phi^-\rangle 
-\frac{\langle \phi^+, \phi^-\rangle}{2g_s\ell_s^5}\langle \phi^+, \Big(
(\bar{z}^i\cdot z^j)^2 -(\bar{z}^i\cdot z^j)(\bar{z}^j\cdot z^i)
\Big)\phi^-\rangle\Big]
|\Psi\rangle .
\end{equation}
These dynamical equations are of course consistent with 
the scaling symmetry which characterizes the space-time uncertainty relation,
\[
(z^i, \bar{z}^i)  \rightarrow \lambda (z^i,\bar{z}^i) \quad t\rightarrow \lambda^{-1}t, 
\quad g_s\rightarrow \lambda^3 g_s
\]
providing that the field operators have scaling property such that 
the combinations $\sqrt{[d^{2d}z]}\phi^{\pm}[z,\bar{z}]$ 
have zero scaling dimension.

\section{Conclusion: a tribute to Yukawa (1907-1981)}
Finally, I would like to give a brief discussion on the nature of 
our D-particle field, in connection with the old idea of `elementary domains' 
by Yukawa. It is well known that Yukawa devoted his research life 
after the meson theory mainly to attempts toward various non-local 
field theories, in hopes of overcoming the ultraviolet difficulty. 
The idea of elementary domains constituted his last work \cite{yukawa} 
along this direction. He proposed to introduce a quantized field 
$\Psi(D, t)$ which is defined on some extended 
spatial domains $D$. He thought that 
various excitations with respect to the domain D would 
give different elementary particles. Of course, this is 
basically the way that we interpret various particle states in string theory. 
In this sense, a general string field $\Psi[x(\sigma)]$ with the 
domain being the one-dimensional string region is quite reminiscent 
of the Yukawa field, if we ignore a difference that his commutation 
relation between creation and annihilation operators deviates from the 
canonical one in some important point. In contrast to this, the string field at 
least in the light-cone gauge obeys the usual canonical commutators 
with respect to the loop space of $x(\sigma)$. 
 It seems also 
that the idea could encompass \cite{stgravity}\cite{shsch} even the 
graviton and hence General Relativity would not have been within his imagination, since he has never discussed gravity 
from his viewpoint of non-local field theories, at least to my knowledge. 

According to his own account, Yukawa's proposal
 was inspired by the famous words by a great Chinese poet Li Po (701-762), saying\footnote{
This is in the preface to ``Holding a Banquet in the Peach and Pear on a 
Spring Night".  The translation adopted in the text here is due to 
{\it Stonebridge Press} (http://www.stonebridge.com/) 
who discusses  the quotations of the same 
sentences made by Basho Matsuo (1644-1694), 
the greatest `Haiku' poet in the Edo period 
in Japan.    
 }

\vspace{0.2cm}
 \begin{center}
 ``{\it  Heaven and earth are the inn
 for all things, \\ the light and shadow the traveler
  of a hundred generations.}"
   \end{center}
\vspace{0.2cm}

\noindent
Yukawa described the inspiration coming to his mind 
 as\cite{yukawa2}
``{\it 
If we replace the words `heaven and earth' by the whole 
3 dimensional space, and `all things' by the elementary 
particles, the 3 dimensional space must consist of 
indivisible minimal volumes, and the elementary 
particles must occupy some of them. 
Such a minimal volume may be called an elementary domain. $\ldots$
}". 

The base space of our D-particle field theory also has some 
resemblance with the idea of elementary domains. In our case, 
the degrees of freedom representing the domains is expressed as 
the infinite dimensional complex vector space of $\{z_n, \bar{z}_n\} \sim D$. Its nature is characterized as follows: 
\begin{itemize}
\item The elementary domains of D-brane fields are represented 
by open-string degrees of freedom and as such  are 
 `fuzzier' than Yukawa's. Our elementary domains are 
a sort of {\sl ``clouds of threads"} emanating from D-particles. 

\begin{figure}[htbp]
\begin{center}
\includegraphics[width=2cm]{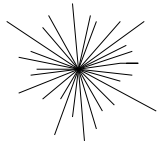}
\caption{\footnotesize{
A D-particle domain as a cloud of open strings.  
}}
\end{center}
\end{figure}

\item The D-brane fields do not satisfy canonical 
commutation relations and are instead characterized by an entirely 
new `continuous' quantum statistics. The domains 
  are mutually permeable and osmotic, in constrast to Yukawa's 
 tighter {\it indivisible} domains. 
\item Our domains are much more dynamical. Infinite components of 
the base-space coorinates are actually latent, and the complexity of domains depends on the number of D-particles. The length scale 
of the domain is governed by  
the space-time uncertainty relation. 
\item The theory includes gravity  through open/closed string duality 
 in the low-energy limit. 
 Even in the non-relativistic approximation, the non-linear 
 general-relativistic  interactions \cite{okyo} of D-particle  can emerge as a quantum loop effect of 
 the Yang-Mills interaction embodied in the field-theory Hamiltonian. 
 So `{\it all things}' could include even space-time geometry in the bulk. 
\end{itemize}

\section*{Acknowledgements}
I would like to thank the organizers 
for invitation to this interesting and enjoyable symposium. 
I have received much inspiration in my youth 
from Yukawa's writings. The present work  is supported in part by Grant-in-Aid No. 16340067 (B))  from the Ministry of  Education, Science and Culture.

%

\end{document}